# Wave propagation in infinite nonlinear acoustic metamaterial beam by considering the third harmonic generation

Xin Fang[1*], Jihong Wen[1#], Dianlong Yu[1], Guoliang Huang[2], Jianfei Yin[1]

[1]*College of Intelligent Science, National University of Defense Technology, Changsha, Hunan 410073, China.*
[2]*Department of Mechanical and Aerospace Engineering, University of Missouri, Columbia, MO 65211, USA*

E-mail: *xinfangdr@sina.com; #jihongwen@vip.sina.com

**Abstract:** Nonlinear acoustic metamaterial (NAM) initiates new fields for controlling elastic waves. In this work, the flexural wave propagation in the half-infinite NAM beam consisting of periodic Duffing resonators is reported by considering the third harmonic generation (THG). Different analytical methods are proposed to describe the wave propagation in the equivalent homogenous medium. Then their effectiveness and accuracy are demonstrated in comparison with the finite element methods. We unveil analytically and numerically extensive physical properties of the strongly nonlinear AM, including the nonlinear resonance in a cell, the effective density, nonlinear locally resonant (NLR) bandgap, propagations and couplings of the fundamental and the third harmonics. These characteristics are highly interrelated, which facilitates the prediction of functionalities. In the near field, the identical bifurcation frequency of these features acts as the start frequency of the NLR bandgap for fundamental waves, whose width is narrower for a stronger nonlinearity. While in the far field, the NLR bandgap characterizes a distance-amplitude-dependent behavior leading to a self-adaptive bandwidth. Moreover, the transmission in the passband of the infinite NAM is different from the chaotic band effect of finite NAMs, and it is influenced by the shifted NLR gap. Our work will promote future studies and constructions of NAMs with novel properties.

## 1. INTRODUCTION

Theories for wave manipulations lay the foundation for advancements in diverse modern techniques. Acoustic metamaterials (AMs) [1-3] and metasurfaces [4-6] provide unconventional functionalities to manipulate subwavelength elastic/acoustic waves such as negative refraction [7], cloaking [8, 9], supper focusing [10], insulation [11-13] and perfect absorption [14, 15]. Lots of them arise from the negative effective mass, $m_{\text{eff}}$, [16, 17] or the negative effective bulk modulus, $B_{\text{eff}}$ [18-20]. For the extensively investigated linear AMs (LAMs) in the past decade, dynamic motions and deformations of the background continuum (such as air or solid) and the embedding structure in the meta-cell follow the linearized laws: Continuity, Hooke's law and Newton's second law [21]. The property of a LAM is unique with specified parameters. However, the propagation wave in AMs becomes nonlinear if there are nonlinearities in the sound field or structures. Rich nonlinear dynamic effects of nonlinear waves and vibrations break new ground for the manipulation of elastic waves.

Nonlinear AM (NAM) is defined as an AM featuring the nonlinear dynamics. Recently, Fang *et al* [22-24] found the chaotic band in finite NAMs and clarified its bifurcation mechanism. Ultra-low and ultra-broad-band wave suppression (the double-ultra effect) is achieved in both one-dimensional and two-dimensional NAMs based on the chaotic bands induced by strongly nonlinear meta-cells [25]. The double-ultra effect breathes new vitality into the field NAM. Bridging coupling of bandgaps can manipulate the bandwidth and the efficiency for the wave reduction in chaotic bands [26]. In addition, multi-state behavior of the wave in the bandgap is observed in NAMs [25]. Bistable periodic structures [27-29] also lead to interesting phenomena such as bandgap transmission.

High-order harmonic generation is an important characteristic of nonlinear systems, and it is active in diverse applications. In nonlinear electromagnetic metamaterials (NEMs) [30], the second- and the third- harmonic generation (SHG and THG) are adopted to realize the phase matching [31, 32]. A combination of the second harmonic in the nonlinear ultrasonic medium with the bandgap of a linear phononic crystal can break the reciprocity at different frequencies [33]. The asymmetric wave propagation under identical frequency is achievable in the nonlinear Schrödinger equation model in theory [34].

Local resonances (LR) in AMs give rise to the wave suppression in the LR bandgap. Passbands of finite LAMs are composed of dense modal resonances of the whole structure. In contrast, the wave propagation in infinite AMs features different process [35] because of the disappearance of boundary reflections, on which occasion resonances of the whole structure are absent. For the proposed AMs made of side holes, Helmholtz resonators or membranes, weak nonlinearities arise when the intensity of the sound field becomes extremely high [36-38]. These nonlinear acoustic fields in infinite AMs lead to the bandgap shifting and the second harmonic generation [39-41]. Perturbation approach is adopted to calculate the effect of the amplitude-dependent small nonlinear correction on the nonlinear bulk modulus [39-41]. For the dispersive wave, Nayfeh and Mook [42] studied the flexural wave propagation in the infinite uniform beam placed on the weakly cubic nonlinear elastic foundation. They considered the interactions between the fundamental and the third harmonics based on the perturbation approach [42].

NAM can boost the development of various methods for achieving tunable, switchable, nonlinear, and sensing functionalities of metamaterials. However, wave propagation in infinite NAMs and its bifurcation mechanisms are not fully understood. Especially under the strongly nonlinear regime, many physical regimes are not clarified, which include: properties of nonlinear effective parameters; the mechanism and distribution of high-order harmonic generation, and its interaction with the fundamental wave; the wave transmissions in bandgaps and passbands; the relationships between the nonlinear resonances, bandgaps, effective parameters and high-order harmonics; and the corresponding mathematical frameworks.

In order to reveal these mechanisms, this paper focuses on the flexural wave propagation in a half-infinite NAM beam consisting of a uniform beam and periodic Duffing resonators. Finite element method and different analytical approaches are proposed to study the properties of the nonlinear effective density $\rho_{\text{eff}}$, nonlinear locally resonant (NLR) bandgap, propagations of the fundamental wave, third harmonic and their couplings. Their relationships and mechanisms arising from bifurcations are elucidated. Moreover, distributions of harmonics, including the new behavior of the nonlinear bandgap, are reported. This work promotes further understandings of NAMs.

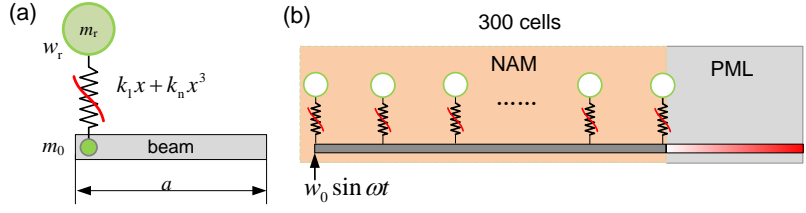

**Figure 1.** The model of the nonlinear acoustic metamaterial beam. (a) A unit cell. (b) The half-infinite finite element model composing of 300 cells and a perfect match layer (PML).

## 2. THEORIES FOR WAVE PROPAGATION

Figure 1 is the schematic of the NAM beam consisting of a uniform beam attached with periodic Duffing oscillators $m_r$ with lattice constant $a$. The density, width and thickness of the primary beam are $\rho$, $b$ and $h$, respectively. $\rho_0=\rho bh$. $m_0$ is the concentrated mass fixed on the beam, and $m_r$ is coupled to $m_0$ through the nonlinear spring $k_r x+k_n x^3$. In theory, the length of the NAM bean is infinite in direction $x$, and a steady incident $w_0=W_0\sin\omega t$ is applied at the boundary $x=0$, where $\omega=2\pi f$ is the angular frequency, $t$ denotes time, and $W_0$ is the constant amplitude. So, it is a half-infinite model. As shown in Figure 1b, a perfect match layer (PML) is built to simulate the infinite boundary in the finite element (FE) model.

### 2.1 The effective mass density

The wave equation of a conventional media is generally written as:

$$D_0 \boldsymbol{L} u + \rho \frac{\partial^2 u}{\partial t^2} = 0 \qquad (1)$$

where $D_0$ is relevant to the stiffness, and $\boldsymbol{L}$ represents the linear partial differential operator. For the flexural wave in a uniform beam, $\boldsymbol{L}=\partial^4/\partial x^4$, $D_0=Ebh^3/12$. In our AM beam, local resonators lead to the negative effective mass density $\rho_{\text{eff}}$. Under the long wave approximation, the wave equation of the linear AM (LAM) beam reads:

$$D_0 \boldsymbol{L} u + \rho_{\text{eff}} \frac{\partial^2 u}{\partial t^2} = 0 \qquad (2)$$

For the LAM beam, substituting the representation $u=Ue^{i(\omega t-kx)}$ gives the solution,

$$\left(k^4-\omega^2 \rho_{\text{eff}}/D_0\right)Ue^{i(\omega t-kx)}=0 \qquad (3)$$

Herein $k$ denotes the wave number. Therefore, the eigen-function reads

$$k^4-\omega^2 \rho_{\text{eff}}/D_0 = 0 \qquad (4)$$

The homogenization approach is adopted to derive the effective parameters. In Figure 1a, $w_0$ and $w_r$ are transverse displacements of $m_0$ and $m_r$ respectively, $u_r=w_r-w_0$, then the motion equation of the Duffing system can be written as:

$$\begin{cases} m_r(\ddot{u}_r+\ddot{w}_0)=-k_r u_r-k_n u_r^3 \\ (m_0+\rho_0 a)\ddot{w}_0+m_r(\ddot{u}_r+\ddot{w}_0)=F(t) \end{cases} \qquad (5)$$

where $F(t)$ is the concentrated force applied on $m_0$. To derive the nonlinear effective density of the unit cell, let's suppose $u_r=U_r\sin\omega t$, $F(t)=F\sin\omega t$. Substituting them into Eq.(5), and adopting the first-order harmonic balance method leads to the algebraic equations:

$$\begin{cases} U_r = \dfrac{\omega^2 m_r W_0}{(k_r - \omega^2 m_r + 3k_n U_r^2/4)} \\ -\omega^2[(m_0 + \rho_0 a)W_0 + m_r(U_r + W_0)] = F \end{cases} \quad (6)$$

A transformation of Eq. (6) gives the effective mass $m_{\text{eff}}$ of the cell.

$$m_{\text{eff}} = \frac{\langle F(t) \rangle}{\langle \ddot{w}_0 \rangle} = \frac{F}{-\omega^2 W_0} = m_0 + \rho_0 a + \frac{k_r + 3k_n U_r^2/4}{\omega_r^2 - \omega^2 + 3k_n U_r^2/(4m_r)} \quad (7)$$

In which $\omega_r = \sqrt{k_r/m_r}$ is the linear nature frequency. $\rho_{\text{eff}} = m_{\text{eff}}/a$. For the linear cell,

$$\rho_{\text{effL}} = \frac{1}{a}\left(m_0 + \rho_0 a + \frac{k_r}{\omega_r^2 - \omega^2}\right) \quad (8)$$

Substituting $\rho_{\text{effL}}$ into Eq. (4) results in the wave number of the LAM beam,

$$k(\omega) = \left[\omega^2 \frac{\rho_{\text{effL}}}{D_0}\right]^{\frac{1}{4}} = \left[\left(\rho_0 a + m_0 + \frac{k_r}{\omega_r^2 - \omega^2}\right)\frac{\omega^2}{D_0 a}\right]^{\frac{1}{4}} \quad (9)$$

With the same approach, the approximate wave number of the NAM model reads

$$k(\omega) = \left[\omega^2 \frac{\rho_{\text{eff}}}{D_0}\right]^{\frac{1}{4}} = \left[\left(\rho_0 a + m_0 + \frac{k_r + 3k_n U_r^2/4}{\omega_r^2 - \omega^2 + 3k_n U_r^2/(4m_r)}\right)\frac{\omega^2}{D_0 a}\right]^{\frac{1}{4}} \quad (10)$$

Therefore, the phase velocity of the flexural wave is expressed as

$$c = \frac{\omega}{k(\omega)} = \omega^{\frac{1}{2}}\left[\left(\rho_0 a + m_0 + \frac{k_r + 3k_n U_r^2/4}{\omega_r^2 - \omega^2 + 3k_n U_r^2/(4m_r)}\right)\frac{1}{D_0 a}\right]^{-\frac{1}{4}} \quad (11)$$

Equation (11) shows the velocity of the wave in NAM is controllable by the amplitude $U_r$.

Unless otherwise mentioned, the parameters specified in simulations are: $m_0=0$, $m_r=0.01$ kg, $\omega_r=80\pi$ rad/s, $k_n=1\times10^9$ N/m$^3$, $b=0.02$ m, $h=0.002$ m, $E=70$ GPa, $\rho_0=2700$ kg/m$^3$, $a=0.04$ m. There are two opposite approaches to solve $U_r$ in Eq. (6): (i) specify the force $F$ but the displacement $W_0$ is unknown; (ii) inversely, specify $W_0$ but $F$ is unknown. Results solved by giving $F=0.2$ N and $W_0=2\times10^{-4}$ m respectively are shown in Figure 2-Figure 4. Nonlinear resonant curves for the two cases are like: there are three branches; branch-2 and branch-3 originate from the saddle-node bifurcation point whose angular frequency is $\omega_J$. The jumping occurs at $\omega_J$. As extensively demonstrated [42], branch-2 corresponds to unstable periodic solutions, i.e., the property depends mainly on branch-1 and branch-3. Though the magnitudes of $U_r$ in the two cases are approximate, $\omega_J$ for $F=0.2$ N is much higher than $\omega_J$ for $W_0=2\times10^{-4}$ m, which indicates that specifying the force has stronger influences on the properties of NAM model.

arXiv: 1808.04682 (14 Aug 2018)

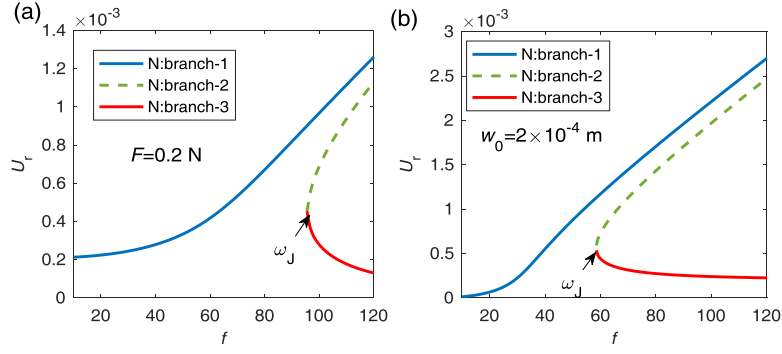

**Figure 2.** Frequency- response curves of the nonlinear resonance in a unit cell. (a) Specify $F=0.2$ N in Eq. (6). (b) Specify $w_0=2\times10^{-4}$ m in Eq. (6). Only real solutions are shown. Dashed curve represents unstable solutions.

Figure 3 illustrates the generalized effective density $\rho_{\text{eff}}/\rho_0$ of the AM model under linear and nonlinear states. The corresponding wave number's real part $\text{Re}(k)$ and the imaginary part $|\text{Im}(k)|$ are shown in Figure 4. For the linear meta-cell, the band for $\rho_{\text{eff}}/\rho_0<0$ keeps constant for specified structural parameters; and the negative $\rho_{\text{eff}}$ gives rise to linear locally resonant (LR) bandgaps whose interval is $(\omega_r, \omega_{cL})$, in which $\text{Im}(k)>0$. $\omega_{cL}$ is the cutoff frequency. Here, the linear LR bandgap is $40<f<73$ Hz, $\omega_{cL}=1.825\omega_r$.

The hardening cubic nonlinearity in the NAM model generates three branches of $\rho_{\text{eff}}$ and $k$ attributing to the bifurcation of $U_r$ in nonlinear resonance (see Figure 2). However, there are great differences between specifying $F$ or $w_0$ in Eq. (6). If $\omega_\infty$ is the frequency for $\rho_{\text{eff}}/\rho_0\to\infty$. $\omega_\infty = \omega_r$ for LAM. As indicated by Figure 3 and Figure 4, on the occasion $F=0.2$ N, the nonlinearity shifts $\omega_\infty$ upward to a higher frequency and a stronger nonlinearity results in a larger deviation to $\omega_r$. Moreover, $\rho_{\text{eff}}/\rho_0$ on branch-1 features unusual behaviors: $\rho_{\text{eff}}/\rho_0$ originates from $-\infty$ near $\omega_\infty$ and $\rho_{\text{eff}}/\rho_0 \to 0$ when $\omega\to\infty$, which indicates the whole branch-1 is a negative density band, and meanwhile $|\text{Im}(k)|>0$ in this range. If $\rho_{\text{eff}}/\rho_0$ jumps to branch-3 at $\omega_J$, though $\rho_{\text{eff}}$ is positive, its $|\text{Im}(k)|$ is still nonzero. Therefore, for the case $\omega>\omega_\infty$, the wave in the NAM model is attenuated.

In contrast, in the situation $W_0=2\times10^{-4}$ m, branch-1 represents $\rho_{\text{eff}}>0$ and $|\text{Im}(k)|=0$; and the negative $\rho_{\text{eff}}$ depends on branch-3. $\omega_J$ initiates the band for $|\text{Im}(k)|>0$, and its cutoff frequency increases to $\omega_{cN}$ in the nonlinear case with $f_J = 58.6$ Hz, $\omega_J =2\pi f_J$. So, the interval $(\omega_J, \omega_{cN})$ denotes the nonlinear LR (NLR) bandgap in the NAM model. As the deviation of $\omega_J$ from $\omega_r$ is much larger than the shifting of the cutoff frequency $\omega_{cL}$ to $\omega_{cN}$, the width of the bandgap becomes narrower in NAMs when specifying the displacement of the fundamental media.

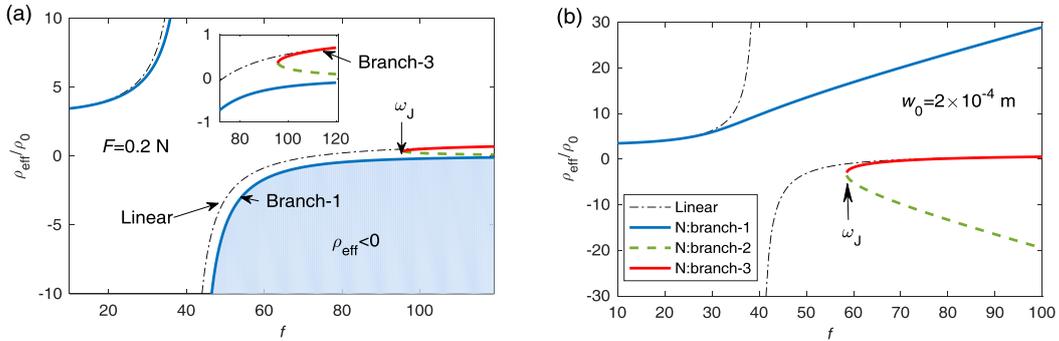

**Figure 3.** The generalized effective density $\rho_{\text{eff}}/\rho_0$ of the acoustic metamaterial beam. The dashed black curves (the thick colorful curves) are linear (nonlinear) results. (a) $F=0.2$ N. (b) $W_0=2\times10^{-4}$ m.

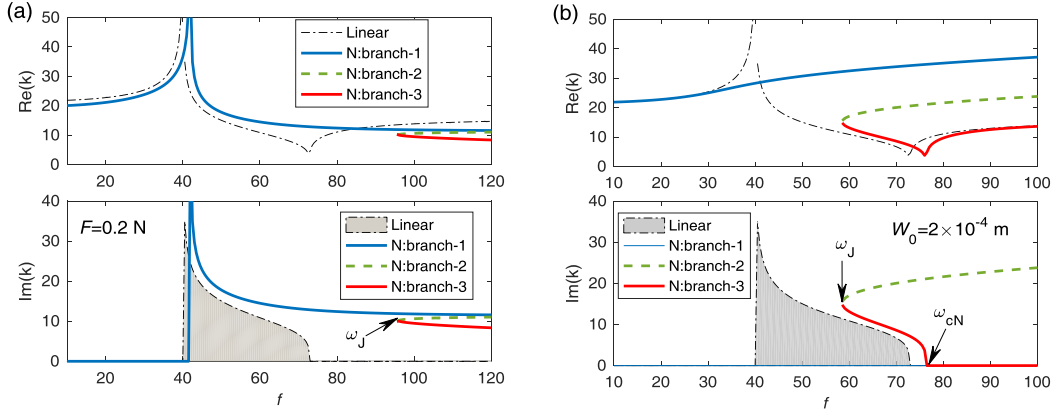

**Figure 4.** Wave number's real part Re(*k*) and the imaginary part |Im(*k*)| corresponding to Figure 3. (a) *F*=0.2 N. (b) $W_0=2\times10^{-4}$ m

### 2.2 The third harmonic generation and interaction

To describe the mechanism for the third harmonic generation (THG) and its coupling with the fundamental wave analytically, the meta-cell is assumed to be equivalent to a homogeneous medium. In our work, the interaction concentrated force *F*(*t*) between the oscillator $m_r$ and the primary beam is equivalent to the uniformly distributive force *f* (*x*, *t*). Moreover, the linear damping effect $\zeta \dot{u}_r$ of $m_r$ is also taken into consideration. Hereby, the coupled wave equation of the NAM beam reads:

$$\begin{aligned}
&\frac{\partial^4 u}{\partial x^4}+\frac{\rho_0}{D_0}\frac{\partial^2 u}{\partial t^2}=-\frac{1}{D_0}f(x,t) \\
&m_r(\ddot{u}_r+\ddot{u})=-\zeta \dot{u}_r - k_r u_r - k_n u_r^3 \\
&m_0 \ddot{u}+m_r(\ddot{u}_r+\ddot{u})=F(t)=af(x,t)
\end{aligned} \quad (12)$$

Moreover, $u_r$ is also a function of the space *x* in the homogenized media, Therefore, Eq. (12) can be simplified as:

$$\begin{cases}
\dfrac{\partial^4 u}{\partial x^4}+\left(\dfrac{\rho_0}{D_0}+\dfrac{m_0+m_r}{D_0 a}\right)\dfrac{\partial^2 u}{\partial t^2}+\dfrac{m_r}{D_0 a}\dfrac{\partial^2 u_r}{\partial t^2}=0 \\
m_r \dfrac{\partial^2}{\partial t^2}(u_r+u)+\zeta \dfrac{\partial u_r}{\partial t}=-k_r u_r - k_n u_r^3
\end{cases} \quad (13)$$

Equation (13) is the coupled flexural wave equation in the equivalent nonlinear homogeneous unconventional media. With considering the third harmonic, the solution of (13) is supposed as the superposition of the fundamental and third harmonics.

$$\begin{cases}
u(x,t)=\dfrac{1}{2}[H_1(\mathbf{x})\exp(i\omega t)+H_1^*(\mathbf{x})\exp(-i\omega t) \\
\qquad\qquad +H_3(\mathbf{x})\exp(3i\omega t)+H_3^*(\mathbf{x})\exp(-3i\omega t)] \\
u_r(x,t)=\dfrac{1}{2}[B_1(\mathbf{x})\exp(i\omega t)+B_1^*(\mathbf{x})\exp(-i\omega t) \\
\qquad\qquad +B_3(\mathbf{x})\exp(3i\omega t)+B_3^*(\mathbf{x})\exp(-3i\omega t)]
\end{cases} \quad (14)$$

in which, *H* and *B* are wave amplitudes; the superscript '*' denotes the conjugation. By substituting Eq. (14) into Eq. (13) and adopting the harmonic balance approach, we yield,

$$\begin{cases} \dfrac{d^4 H_1}{dx^4} - \left(\dfrac{\rho_0}{D_0} + \dfrac{m_0 + m_r}{D_0 a}\right)\omega^2 H_1 - \dfrac{m_r}{D_0 a}\omega^2 B_1 = 0 \\ \dfrac{d^4 H_3}{dx^4} - 9\left(\dfrac{\rho_0}{D_0} + \dfrac{m_0 + m_r}{D_0 a}\right)\omega^2 H_3 - \dfrac{9 m_r}{D_0 a}\omega^2 B_3 = 0 \\ \omega^2 m_r (H_1 + B_1) = i\zeta\omega B_1 + k_r B_1 + \dfrac{1}{4} k_n (3 B_1^2 B_1^* + 3 B_1^{*2} B_3 + 6 B_1 B_3 B_3^*) \\ 9\omega^2 m_r (H_3 + B_3) = 3 i\zeta\omega B_3 + k_r B_3 + \dfrac{1}{4} k_n (B_1^3 + 6 B_1 B_1^* B_3 + 3 B_3^2 B_3^*) \end{cases} \quad (15)$$

The coupling terms in (15) lead to complicate solutions. To describe the property of the NAM beam, we derive analytical solutions in different situations: unidirectional coupled (UC) solution, weakly nonlinear (WN) solution, and fully coupled (FC) solution.

*A. Unidirectional coupled solution*

To obtain succinct but accurate solution of Eq.(15), we firstly neglect the influence of the third harmonic on the fundamental wave, but consider the energy transformation from the fundamental wave into its third harmonic, i.e., the unidirectional coupling (UC). Then, one obtains equations relevant to the fundamental wave only:

$$\begin{cases} \dfrac{d^4 H_1}{dx^4} - \left(\dfrac{\rho_0}{D_0} + \dfrac{m_0 + m_r}{D_0 a}\right)\omega^2 H_1 - \dfrac{m_r}{D_0 a}\omega^2 B_1 = 0 \\ \omega^2 m_r (H_1 + B_1) = i\zeta\omega B_1 + k_r B_1 + \dfrac{3}{4} k_n B_1^2 B_1^* \end{cases} \quad (16)$$

Supposing the solution of Eq. (16) takes the form:

$$\begin{aligned} H_1(x) &= h_1 \exp(-i k_1 x), & H_1^*(x) &= h_1^* \exp(i k_1 x) \\ B_1(x) &= b_1 \exp(-i k_1 x), & B_1^*(x) &= b_1^* \exp(i k_1 x) \end{aligned} \quad (17)$$

Herein $k_1$ denotes the wave number of the fundamental wave; $h_1$ and $b_1$ are amplitudes. Substituting Eq. (17) into Eq. (16) leads to

$$\begin{cases} k_1^4 h_1 - \left(\dfrac{\rho_0}{D_0} + \dfrac{m_0 + m_r}{D_0 a}\right)\omega^2 h_1 - \dfrac{m_r}{D_0 a}\omega^2 b_1 = 0 \\ \omega^2 m_r (h_1 + b_1) = i\zeta\omega b_1 + k_r b_1 + \dfrac{3}{4} k_n b_1^2 b_1^* \end{cases} \quad (18)$$

Equation (18) can be evaluated to yield

$$k_1^4 = \dfrac{\omega^2}{D_0 a}\left(\rho_0 a + m_0 + \dfrac{k_r + 3 k_n b_1 b_1^* / 4 + i\zeta\omega}{\omega_r^2 - \omega^2 + (3 k_n b_1 b_1^* / 4 + i\zeta\omega)/m_r}\right) \quad (19)$$

In fact, if $\zeta=0$, Eq. (19) takes the same form with Eq. (10); furthermore, with a given $h_1$ in Eq. (18), the analytical result derived with Eq. (19) should be equal to Eq. (10) in the situation of specifying the displacement $W_0 = h_1$ in Eq. (6).

If the damping is taken into consideration, by decomposing $h_1$ and $b_1$ into the real and imaginary parts $b_1 = b_{1r} + i b_{1I}$, $h_1 = h_{1r} + i h_{1I}$, one can factorize the second formula in Eq. (18):

$$\begin{cases} \omega^2 m_r (h_{1r} + b_{1r}) = -\zeta\omega b_{1I} + k_r b_{1r} + \dfrac{3}{4} k_n (b_{1r}^2 + b_{1I}^2) b_{1r} \\ \omega^2 m_r (h_{1I} + b_{1I}) = \zeta\omega b_{1r} + k_r b_{1I} + \dfrac{3}{4} k_n (b_{1r}^2 + b_{1I}^2) b_{1I} \end{cases} \quad (20)$$

When the third harmonic propagating along the NAM beam, the nonlinear effect arising from itself is neglectable if the nonlinearity is not extremely strong. On this occasion, we obtain equations about the third harmonic by removing the nonlinear terms about $B_3$ in Eq. (15), i.e.,

$$\begin{cases} \dfrac{d^4 H_3}{dx^4} - 9\left(\dfrac{\rho_0}{D_0} + \dfrac{m_0 + m_r}{D_0 a}\right)\omega^2 H_3 - 9\dfrac{m_r}{D_0 a}\omega^2 B_3 = 0 \\ 9\omega^2 m_r (H_3 + B_3) = 3i\zeta\omega B_3 + (k_r + \dfrac{3}{2}k_n b_1 b_1^*)B_3 + \dfrac{1}{4}k_n b_1^3 e^{-i3k_1 x} \end{cases} \quad (21)$$

Because $b_1$ is solvable from Eq. (20), Eq. (21) becomes a system of linear differential equations with the forward-propagating driving term $k_n b_1^3 e^{-i3k_1 x}/4$. This is also the process for THG. The solution is thereby a superposition of the particular and general solutions, i.e.,

$$\begin{aligned} H_3(x) &= h_{31} \exp(-3ik_1 x) + h_{32} \exp(-ik_3 x) \\ B_3(x) &= b_{31} \exp(-3ik_1 x) + b_{32} \exp(-ik_3 x) \end{aligned} \quad (22)$$

Amplitudes $h_{3j}$ and $b_{3j}$ are determined by initial conditions. Because the third harmonic is absent at the boundary $x=0$, one yields $h_{32} = -h_{31}$. Thus, there are at least two wave modes in the third harmonic whose wave numbers are $3k_1$ and $k_3$, respectively. For nondispersive media and nondispersive waves, $3k_1=k_3$, so as to compose the two modes together. However, AMs are dispersive media and flexural waves are also dispersive, we cannot directly specify $3k_1=k_3$. Substituting Eq. (22) into Eq. (21) results in:

$$\begin{cases} h_{31}(3k_1)^4 - 9\left(\dfrac{\rho_0}{D_0} + \dfrac{m_0 + m_r}{D_0 a}\right)\omega^2 h_{31} - \dfrac{9m_r}{D_0 a}\omega^2 b_{31} = 0 \\ 9\omega^2 m_r (h_{31} + b_{31}) = (3i\zeta\omega + k_r + \dfrac{3}{2}k_n b_1 b_1^*)b_{31} + \dfrac{1}{4}k_n b_1^3 \end{cases} \quad (23)$$

$$\begin{cases} h_{32} k_3^4 - 9\left(\dfrac{\rho_0}{D_0} + \dfrac{m_0 + m_r}{D_0 a}\right)\omega^2 h_{32} - \dfrac{9m_r}{D_0 a}\omega^2 b_{32} = 0 \\ 9\omega^2 m_r (h_{32} + b_{32}) = (3i\zeta\omega + k_r + \dfrac{3}{2}k_n b_1 b_1^*)b_{32} \end{cases} \quad (24)$$

Equation (24) gives us an Eigen-function:

$$k_3^4 - \dfrac{9\omega^2}{D_0 a}(\rho_0 a + m_0 + m_r) - \dfrac{9m_r \omega^2}{D_0 a}\dfrac{9m_r \omega^2}{(k_r + 3i\zeta\omega + 3k_n b_1 b_1^*/2 - 9\omega^2 m_r)} = 0 \quad (25)$$

The solution of $k_3$ with nonzero real part is

$$k_3 = \left(\rho_0 a + m_0 + \dfrac{k_r + 3i\zeta\omega + 3k_n b_1 b_1^*/2}{\omega_r^2 - 9\omega^2 + (3i\zeta\omega + 3k_n b_1 b_1^*/2)/m_r}\right)^{1/4} \left(\dfrac{9\omega^2}{D_0 a}\right)^{1/4} \quad (26)$$

This formula indicates that $k_3$ depends on the amplitude of the fundamental wave $b_1$. Furthermore, a conversion of Eq. (23) leads to the explicit solution:

$$\begin{aligned} h_{31} &= \dfrac{-1}{(3k_1)^4 - k_3^4}\dfrac{9\omega^2}{D_0 a}\dfrac{k_n b_1^3 / 4}{[\omega_r^2 - 9\omega^2 + (3i\zeta\omega + 3k_n b_1 b_1^*/2)/m_r]} \\ b_{31} &= \dfrac{9\omega^2 m_r h_{31} - k_n b_1^3/4}{k_r - 9\omega^2 m_r + 3i\zeta\omega + 3k_n b_1 b_1^*/2} \end{aligned} \quad (27)$$

Moreover, we know from Eqs. (26) and (19) that $k_3(\omega) \neq k_1(3\omega)$ as the deferent coefficient of the nonlinear term $k_n b_1 b_1^*$, but there is $k_3(\omega) \approx k_1(3\omega)$ if $|b_1|$ and $k_n$ are small.

*B. Weakly nonlinear solution*

When the nonlinearity is weak, the nonlinear terms relevant to the fundamental wave in Eq. (15) are also neglectable. In this case, the influences of the fundamental wave on its third harmonic are still considered. Thus, one yields the weakly nonlinear (WN) solution:

$$\begin{cases} b_1 = \dfrac{\omega^2}{\omega_r^2 - \omega^2 + i\zeta\omega/m_r} h_1, \\ k_1 = \left[\left(\rho_0 a + m_0 + \dfrac{k_r + i\zeta\omega}{\omega_r^2 - \omega^2 + i\zeta\omega/m_r}\right)\dfrac{\omega^2}{D_0 a}\right]^{1/4}, \\ k_3 = \left[\left(\rho_0 a + m_0 + \dfrac{k_r + 3i\zeta\omega}{\omega_r^2 - 9\omega^2 + 3i\zeta\omega/m_r}\right)\dfrac{9\omega^2}{D_0 a}\right]^{1/4}, \\ h_{30} = \dfrac{-1}{(3k_1)^4 - k_3^4}\dfrac{9\omega^2}{D_0 a}\dfrac{k_n b_1^3/4}{\omega_r^2 - 9\omega^2 + 3i\zeta\omega/m_r} \end{cases} \quad (28)$$

In WN solution, linear effect is considered only, so $k_1(3\omega) = k_3(\omega)$. If $\zeta=0$, $h_{30}$ characterizes two peaks at $\omega_r/3$ and the frequency for $3k_1=k_3$ respectively. Solving $3k_1=k_3$ leads to two frequency solutions:

$$\omega = \dfrac{\omega_r}{3}\sqrt{5\gamma_m \pm \sqrt{25\gamma_m^2 - 9\gamma_m}}, \quad \gamma_m = 1 + \dfrac{m_r}{\rho_0 a + m_0} \quad (29)$$

This equation indicates that the highly efficient energy generation in weakly nonlinear case depends on the attached mass ratio. If $\gamma_m \to 1$, the two solutions are $\omega \to \omega_r$ and $\omega \to \omega_r/3$. If $\gamma_m = 1.5$, they are $\omega = 0.327\omega_r$, $\omega = 1.249\omega_r$.

Moreover, if the perfect phase matching is anticipated in this case, the two solutions for $3k_1=k_3$ should just right be $\omega$ and $3\omega$, i.e., $\gamma_m^2 - \gamma_m = 0$, so $\gamma_m = 1$. However, there must be $\gamma_m > 1$ in metamaterials, so the perfect phase matching in this case is unrealizable. If the nonlinearity is not weak, the peak frequency of $h_{30}$ is not only depends on the attached mass ratio, but also on the amplitude.

The UC approach neglects influences of the third harmonic on the fundamental wave, and the nonlinear effects arising from the third harmonic itself. To consider all these couplings, this work derives the fully coupled (FC) solution.

*C. Fully coupled solution*

To obtain FC solution, the damping is not considered. Based on the analyses above, the solution of the differential equation (13) is supposed as:

$$\begin{aligned} u &= h_1 \sin(\omega t - k_1 x) + h_{31} \sin(3\omega t - k_3 x) + h_{32} \sin(3\omega t - 3k_1 x) \\ u_r &= b_1 \sin(\omega t - k_1 x) + b_{31} \sin(3\omega t - k_3 x) + b_{32} \sin(3\omega t - 3k_1 x) \end{aligned} \quad (30)$$

Similarly, with the boundary condition at $x=0$, one yields $h_{32} = -h_{31}$. The expression $u_r^3$ is reduced as

$$\begin{aligned} u_r^3 =& \dfrac{3}{4}\left(b_1^3 + 2b_1 b_{31}^2 + b_1^2 b_{32} + 2b_1 b_{32}^2\right)\sin(\omega t - k_1 x) + \\ & \dfrac{3}{4}\left(2b_1 b_{31} b_{32} - b_1^2 b_{31}\right)\sin[\omega t - (k_3 - 2k_1)x] + \dfrac{3}{2} b_1 b_{31} b_{32} \sin[\omega t - (4k_1 - k_3)x] \\ & + \dfrac{3}{4}\left(2b_1^2 b_{31} + 2b_{31} b_{32}^2 + b_{31}^3\right)\sin(3\omega t - k_3 x) + \\ & \left(-\dfrac{1}{4}b_1^3 + \dfrac{3}{2}b_1^2 b_{32} + \dfrac{3}{2}b_{31}^2 b_{32} + \dfrac{3}{4}b_{32}^3\right)\sin(3\omega t - 3k_1 x) \\ & + \dfrac{3}{4} b_{31}^2 b_{32} \sin[3\omega t - (2k_3 - 3k_1)x] + \dfrac{3}{4} b_{31} b_{32}^2 \sin[3\omega t - (6k_1 - k_3)x] + O(\omega^5) \end{aligned} \quad (31)$$

Besides the modes $k_1$, $k_3$ and $3k_1$ introduced in Eq. (30), the first-order modes $k_3$-$2k_1$ and $4k_1$-$k_3$,

along with the third-order modes $2k_3-3k_1$ and $6k_1-k_3$ appear in Eq. (31). If $k_3=3k_1$, these modes will be reduced to $k_1$ and $k_3$. This manuscript considers the more general case $k_3 \neq 3k_1$ in the strongly dispersive medium. With a substitution Eq. (30) into Eq. (13), and adopting the harmonic balances of three modes $k_1$, $k_3$ and $3k_1$, one yields a system of algebraic equations for FC solution:

$$\begin{cases} k_1^4 h_1 - \dfrac{\omega^2}{D_0 a}(\rho_0 a + m_0 + m_r) h_1 - \dfrac{m_r}{D_0 a}\omega^2 b_1 = 0 \\ k_3^4 h_{31} - \dfrac{9\omega^2}{D_0 a}(\rho_0 a + m_0 + m_r) h_{31} - \dfrac{9 m_r}{D_0 a}\omega^2 b_{31} = 0 \\ (3k_1)^4 h_{32} - \dfrac{9\omega^2}{D_0 a}(\rho_0 a + m_0 + m_r) h_{32} - \dfrac{9 m_r}{D_0 a}\omega^2 b_{32} = 0 \\ \omega^2 m_r (b_1 + h_1) = k_r b_1 + k_n (\dfrac{3}{4}b_1^3 + \dfrac{3}{2}b_1 b_{31}^2 + \dfrac{3}{4}b_1^2 b_{32} + \dfrac{3}{2}b_1 b_{32}^2) \\ 9\omega^2 m_r (b_{31} + h_{31}) = k_r b_{31} + k_n (\dfrac{3}{2}b_1^2 b_{31} + \dfrac{3}{2}b_{31} b_{32}^2 + \dfrac{3}{4}b_{31}^3) \\ 9\omega^2 m_r (b_{32} + h_{32}) = k_r b_{32} + k_n (-\dfrac{1}{4}b_1^3 + \dfrac{3}{2}b_1^2 b_{32} + \dfrac{3}{2}b_{31}^2 b_{32} + \dfrac{3}{4}b_{32}^3) \end{cases} \quad (32)$$

By combining the condition $h_{32}=-h_{31}$, six unknown parameters are solvable by specifying $h_1$.

## 3. SIMULATIONS AND ANALYSES

To demonstrate the analytical methods and investigate properties of the wave propagation in the half-infinite NAM beam, the COMSOL finite element (FE) model composed of 300 unit cells and a perfect match layer (PML) is established, as shown in Figure 1b. The PML consists of 10 segments of uniform beams with gradually increasing structural damping. The length of each segment is $20a$, and its structure parameters are identical with the primary beam in NAM. Rayleigh damping $\xi_i=\beta_i K$ is adopted to attenuate the wave in PML, where $K$ denotes the stiffness of the beam. The stiffness damping factor of the $i$th segment is $\beta_i = c_{max}(i/N_p)^3$, $c_{max}=1$, $N_p=10$. The designed PML simulates the nonreflecting boundary condition accurately. Moreover, a finite FE model consisting of 1500 cells is also established for necessary comparison. In FE models, the incident transverse wave at the boundary $x=0$ is $W_0 \sin(2\pi f_F t)$, $W_0 = 2 \times 10^{-4}$ m, and $f_F$ is the frequency of the fundamental wave. This is a displacement boundary. We note that, though the force is used in calculating the effective mass density, it cannot be used in time-domain simulations, on which occasion the response of the free structure approaches to infinite. Unless otherwise specified, other parameters are same with Figure 2.

Typical profiles of the wave propagation in the NAM beam are shown in Figure 5. In time domain, though the incident wave is steady, the profiles of the propagating waves are not steady. The analyzed time interval at the $n^{th}$ cell is ($t_{st, n}$, $t_{ct, n}$). We keep $\Delta t = t_{ct, n} - t_{st, n}$ of different points equal but the starting time $t_{st, n}$ increases with the propagation distance $na$.

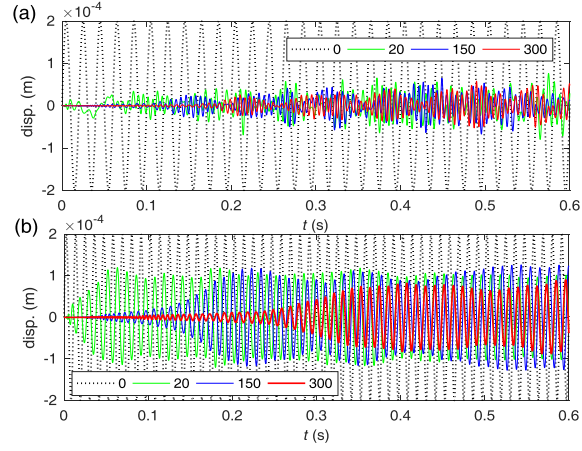

**Figure 5.** Profile of the wave in the NAM beam. (a) $f_F = 50$Hz in the linear bandgap; (b) $f_F = 90$ Hz in the passband. The number, $n$, in legends denotes the propagation distance is $x=na$.

### 3.1 Frequency components in harmonics

Frequency components in harmonics should be clarified in advance of analyzing the property of wave propagation. Figure 6 shows typical spectra of waves in the passband ($f_F = 35$ Hz) and LR bandgap ($f_F = 50$ Hz). As indicated by the frequency spectrum $|P(f)|$, high-order harmonics become remarkable as the propagation distance $na$ increases; moreover, the third and fifth harmonics are main components among high-order waves. However, they do not always localize at the frequencies $3f_F$ and $5f_F$, but also at $3.5f_F$ and $5.5f_F$ instead. Therefore, the peak amplitude of $|P(f)|$ in the interval $2.5f_F$-$3.5f_F$ is regarded as the amplitude of the third harmonic $A_3$. $A_t$ is maximum amplitude of the wave in the analyzed time interval. $A_1$ is the amplitude of the fundamental wave. $A_p$ denotes the peak amplitude of $|P(f)|$, and $f_p$ is the corresponding frequency. The distribution of the generalized peak frequency $\Omega=f_p/f_F$ against $na$ and $f_F$ is shown in Figure 7. For the analyzed nonlinear strength, results indicate that: fundamental wave is generally the main component for $f_F$ in the passbands; in the frequency range for highly efficient THG (we will detail it later), $\Omega=3$~$3.5$; but in the range 60-70 Hz insides the LR bandgap, the second harmonic is also obvious.

Both $f_F=35$ Hz and $f_F=50$ Hz come from the band for efficient THG. As illustrated in Figure 6b,d, there is $A_p \geq A_1$ for both linear and nonlinear AMs; $A_3$ keeps almost steady for $n>100$, and $A_p=A_3$ after certain distance $n_c$ where the third harmonic becomes the main component of the wave. Because of the bandgap effects on the fundamental wave, $n_c = 10$ for $f_F = 50$ Hz. But $n_c = 290$ for $f_F=35$ Hz.

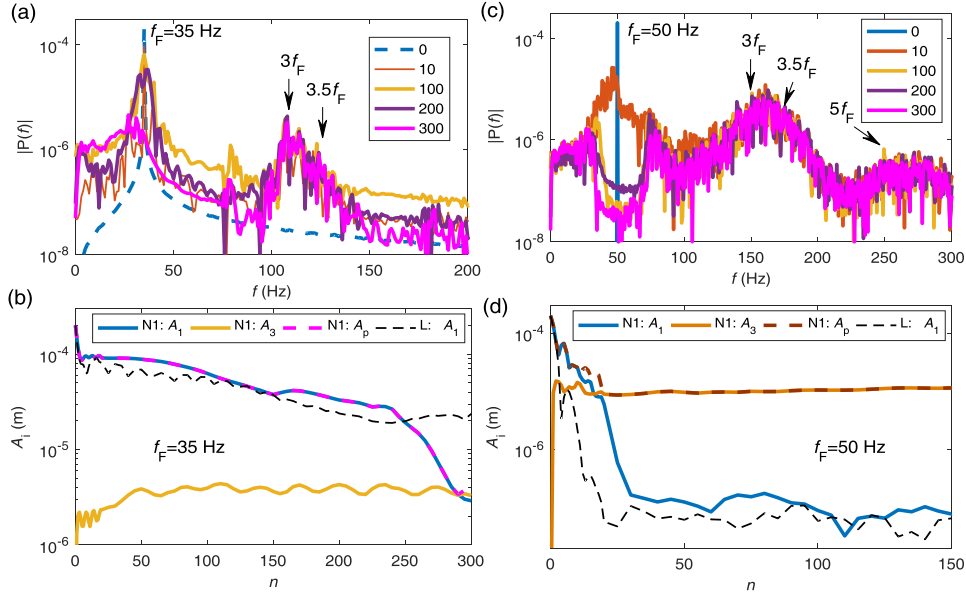

**Figure 6.** Frequency spectrum $|P(f)|$ and the amplitudes $A_i$ of the main components for the wave at $n$th cell. (a, b) $f_F = 35$ Hz in the first passband. (c, d) $f_F = 50$ Hz inside the bandgap. In the legends of panels (a, c), the number $n$ denotes the coordinate is $x=na$. In the panels (b, d), N1: $k_n$=1e9; L: the linear case $k_n$=0.

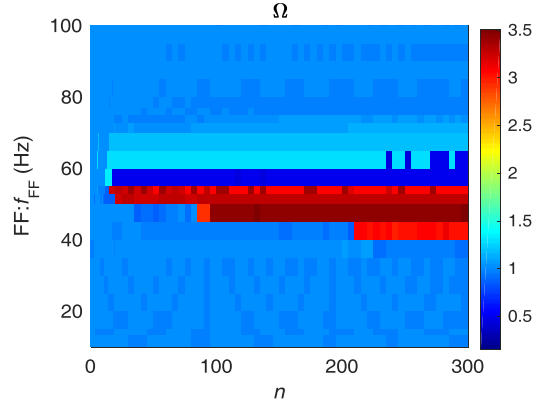

**Figure 7.** Distribution of the generalized peak frequency $\Omega=f_p/f_F$.

### 3.2 Fundamental wave, TGH and their interactions

This section studies properties of the fundamental and the third harmonics by combining the analytical solutions with the numerical method without the damping effect. In analytical approaches, real solutions are solved and the input amplitude of the fundamental wave is $h_1=W_0=0.2$ mm, which is equal to the amplitude for deriving $\rho_{\text{eff}}$. In analytical solutions, $h_3=|h_{31}|$. Figure 8 shows the relationship $h_3$-$f_F$, and the corresponding wave numbers $k_1$ and $k_3$ are shown in Figure 9.

Only linear effects of the fundamental and third harmonics are considered in the WN model, so that $k_1(3\omega) = k_3(\omega)$ and its curves feature no bifurcations. This model anticipates that THG has two peaks at $\omega_F=\omega_r/3$ and $\omega_F\approx\omega_r$ respectively. Near $\omega_F\approx\omega_r$, the incident wave gives rise to the fundamental nonlinear resonances of the oscillators $m_r$ that generate third harmonic efficiently. At $\omega_F=\omega_r/3$, THG arises from the third super-harmonic resonances of $m_r$. However, the efficiency for THG near $\omega_r/3$ is much lower than the band near $\omega_r$ in theory. As indicated by Im($k_j$) in Figure 9a, the third harmonic $3\omega_F$ with wave number $3k_1$ is stopped for $\omega_r< \omega_F <\omega_{cL}$ (the linear LR bandgap); and $3\omega_F$ with wave number $k_3$ is stopped if $\omega_r<3\omega_F<\omega_{cL}$. Therefore, the range $\omega_r/3<\omega_F<0.61\omega_r$ (i.e., 13.5<$f_F$<24.3Hz here) is a valley interval for $h_3$ (see Figure 8a).

There are three branches of THG solved with UC model in Figure 8a. Because $W_0=h_1$ is specified in theory, the fundamental wave number derived with $\rho_{eff}$ (described by Eq. (10)) is equal to $k_1$ solved with UC model. Curves in Figure 9a confirm this fact. Therefore, there is a one-to-one correspondence between each branch of $h_3$ in UC solutions and each branch of the nonlinear resonance illustrated in Figure 2b, so that they get identical bifurcation frequencies $\omega_J$ (58.6 Hz for the specified parameters). Accordingly, the nonlinear resonance in a unit cell is helpful to explain the properties for THG in the whole NAM. The UC model still predicts the peak of $h_3$ near $\omega_F=\omega_r/3$, and its valley region $\omega_r/3<\omega_F<0.61\omega_r$ agrees with the WN result well. However, there is not an obvious peak near $\omega_r$ on branch-1 of UC solution. Instead, branch-1 gradually increases to a saturated value after the valley. According to the nonlinear resonance, branch-2 is unstable solution, thus branch-3 may dominate the characteristic of THG for $\omega_F>\omega_J$. $h_3$ may jump from branch-1 to branch-3 at $\omega_J$ leading to the reduction of the efficiency for THG. Moreover, it is surprising that a peak of $h_3$ appear on branch-3. As indicated by Im($k_1$), the band $\omega_J<\omega_F<\omega_{cN}$ is the NLR gap of the third harmonic with wave number $3k_1$ and the fundamental wave. Im($k_3$) presented by UC solution is coincide with the WN solution here.

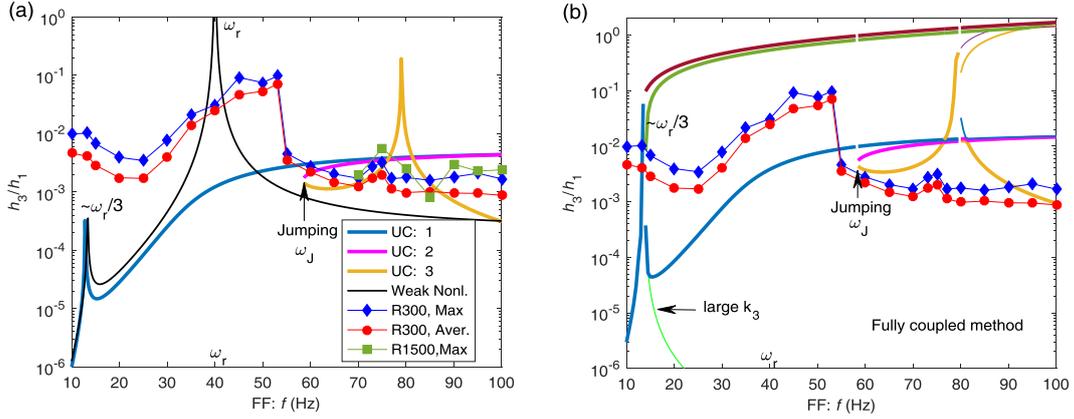

**Figure 8.** Third-harmonic amplitude $h_3$ varying against the fundamental frequency (FF) $f_F$ solved with different methods. (a) Unidirectional coupled (UC) solution and weakly nonlinear solution; (b) Fully coupled solution, $h_3=|h_{31}|$. In (a), UC: $j$ labels the $j$th branch. The dotted curves delineate numerical results from the FE model, $h_3=A_3$. In legends, "R300, Max" represents the maximum $A_3$ for cells 2<$n$<300 in the infinite FE model consisting of 300 cells and PML; "R300, Aver." is the average $A_3$ in that space; "R1500, Max" represents maximum $A_3$ for 2<$n$<300 in the finite FE model consisting of 1500 cells. $\omega_J$ denotes the saddle-node bifurcation point.

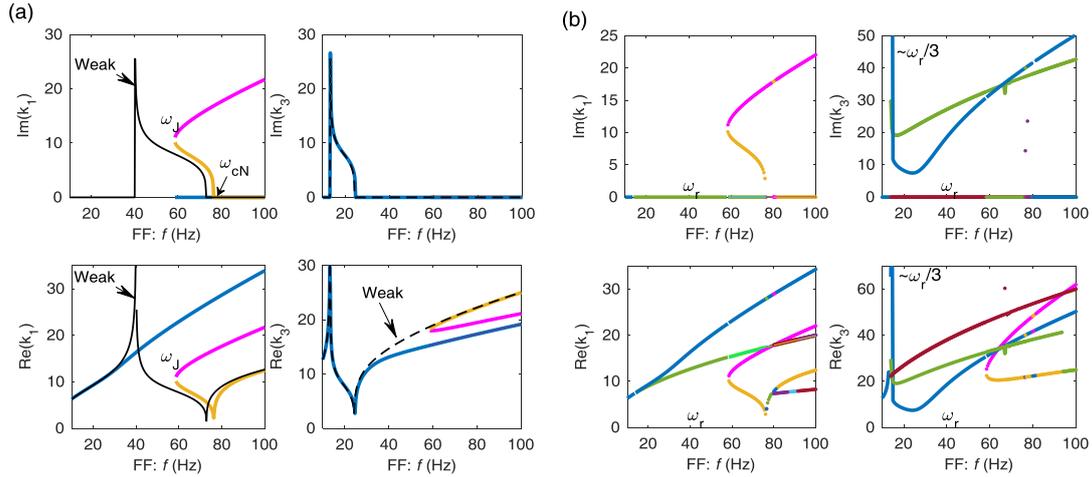

**Figure 9.** Wave numbers $k_1$ and $k_3$ solved with different method. (a) UC and weakly nonlinear solutions. (b) Fully coupled solution. Different curves are in accordance with different branches in Figure 8.

Before the statements of the FC solution, we describe the numerical results based on FE models, as shown in Figure 8 the discrete points. There are unavoidable numerical errors arising from the Fast Fourier Transform when calculating a spectrum to obtain $A_1$ and $A_3$ in FE models. Although there are significant differences between the numerical results and UC solutions, their tendencies, jumping frequency and peaks of curves are consistent. FE results verify the peak near $\omega_r/3$ and the valley in 14-30 Hz. For $f_F > 30$ Hz, $h_3$ gradually increases to a saturate value and features a highly efficient THG band in 35 <$f_F$ <53 Hz. However, there is not an obvious peak as predicted by the WN solution, so WN model fails in describing the features behind the valley. At 53 Hz, $h_3$ jumps suddenly to a low-amplitude trajectory, which reproduces the bifurcation frequency $\omega_J$ accurately. Therefore, $\omega_J$ does not only initiate the NLR gap of fundamental waves; it is also the cutoff frequency for highly efficient THG near $\omega_r$. Moreover, numerical results on this trajectory follow the same regular as branch-3 in UC solutions, including the peak near 75 Hz, which demonstrates that branch-3 dominates the property after the jump. To verify further the peak near 75 Hz is the essential characteristic of the NAM, we calculate the response of the FE model consisting of 1500 cells and analyze the signal in the space $a<na<300a$. These results confirm this peak again. However, its amplitude is much lower than the prediction of UC solution. Numerical analyses demonstrate the validity of UC solutions in predicting the features of the wave propagation and THG.

FC solution considers more interactions. As shown in Figure 8b and Figure 9b, FC solution characterizes complicate bifurcations. There are three backbone curves following the same laws as UC solution, including the bifurcation, the valley and peaks. We also labeled them as branch-1/2/3 correspondingly. However, FC model presents higher $h_3$ that is closer to numerical results. Except for other multiple branches, properties of $k_1$ indicated by branch-1/2/3 in UC and FC solutions are approximately equal. On branch-3, they predict the propagating third harmonic's mode is $k_3$. However, behaviors of $k_3$ on branch-1 are different. For $\omega_F > \omega_r/3$, the whole branch-1 in FC solution corresponds to Im($k_3$)>0 leading to the attenuation of the third harmonic with wave number $k_3$. In the band 35-53 Hz for highly efficient THG, both modes $3k_1$ and $k_3$ for the third harmonics can propagate. Moreover, there are multiple branches in FC solutions, in which two large-amplitude branches originate from the bifurcation near $\omega_r/3$, but the low-amplitude branch has $k_3$>600 indicating slow waves or attenuated waves. Generally speaking, the FC model fits the numerical result better, but its complex bifurcations make the analyses difficult.

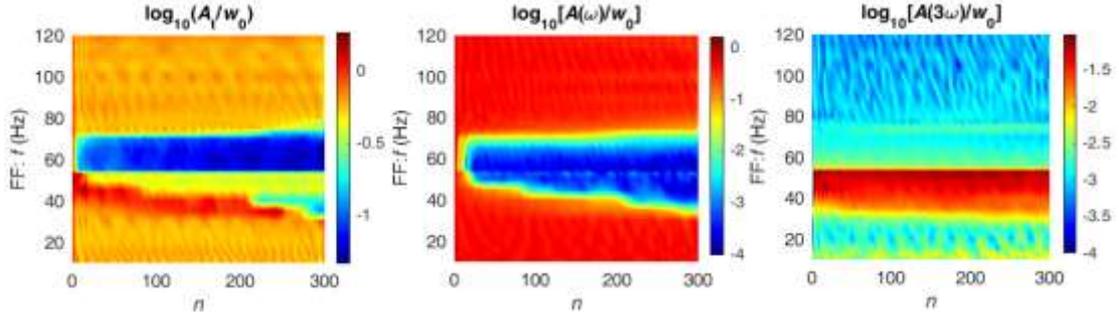

**Figure 10.** Different generalized amplitudes varying with the propagation distance *n* and $f_F$ in FE model. $A(\omega)=A_1$, $A(\omega)=A_3$.

Figure 10 illustrates the distributions of generalized amplitudes $\log_{10}(A_i/W_0)$ against the propagation distance *n* and $f_F$ presented by the half-infinite FE model. The energy is convert to the third harmonic efficiently in the band 35< $f_F$ <53 Hz. In most cases, as the propagation distance *n* increases, $A_3$ keeps steady expect for inevitable fluctuations, i.e., the vast majority of the third harmonic generates in the near field. There are exceptional frequencies $f_F$ < 20 Hz whose $A_3$ increases with the propagation distance.

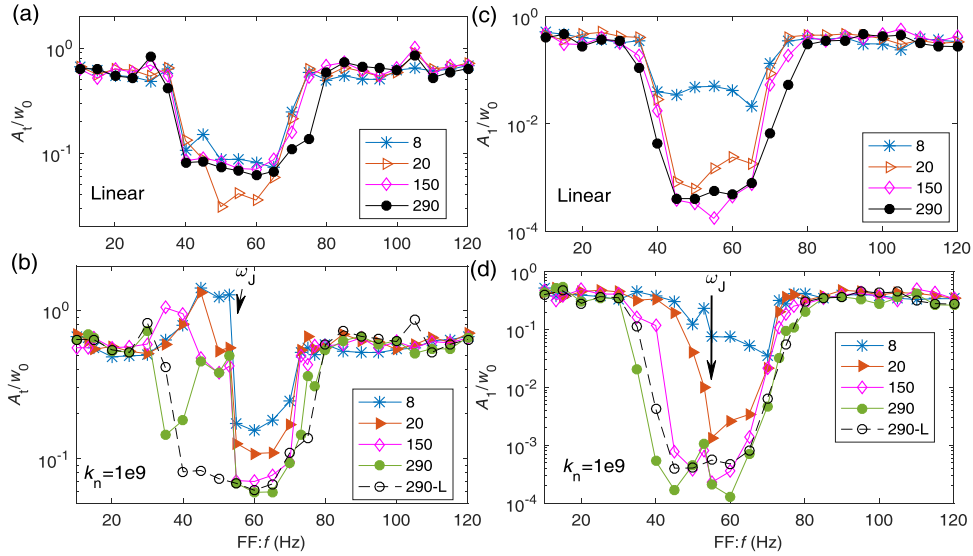

**Figure 11.** Varying laws of the total amplitude $A_t$ and the fundamental wave's amplitude $A_1$ at the *n*th cell. (a, b) $A_t/w_0$; (c, d) $A_1/w_0$. (a, c) linear results; (b, d) comparisons between linear and nonlinear results. In legends, numbers denote the *n*th cell, and "290-L" represents the 290th cell in the linear model.

In contrast, significant changes of $A_t$ and $A_1$ are observed as *n* increases. More details are presented in Figure 11. For the LAM model, $A_t$ and $A_1$ in the bandgap keep almost steady for *n*>20. For the specified team of parameters, one obtains $\omega_J < \omega_{cL}$ for NAM beam. In the near field when *n*<20, the NAM generates the attenuation effect of fundamental waves in the band $\omega_J<\omega_F<\omega_{cL}$ only, indicating numerically that $\omega_J$ initiates the near-field NLR bandgap. In this adjustable gap, both the fundamental and third harmonics are stopped. However, the NLR bandgap does not keep constant as the distance increases, instead, $A_1$ in $\omega_r< \omega_F <\omega_J$ reduces with the propagation distances, which results in the range ($\omega_r$, $\omega_{cL}$) still become the bandgap in the far field. The origin of this behavior is due to the amplitude-dependent property of the dispersion relation. In $\omega_r < \omega_F < \omega_J$, as the propagation distance increases, the attenuation of the fundamental wave weakens the nonlinear strength, then a weaker nonlinearity actually leads to a lower bifurcation frequency $\omega_J$; conversely, the lower $\omega_J$ reduces $A_1$ further. This process appears in turn, leading to the far-field $\omega_J$ approaching to the start frequency of linear bandgap $\omega_r$. In short, $\omega_J$ depends on the fundamental wave's amplitude and the amplitude depends on the propagation distance. Such a distance-amplitude-dependent property of $\omega_J$ results in highly self-adaptive behavior of the NLR gap's width to the propagation distance. However, in $\omega_r < \omega_F < \omega_J$, $A_t$ does not reduce as much as $A_1$ owing to the high-amplitude third harmonics. Therefore, the near-field $\omega_J$ is still the start frequency of the far-field nonlinear bandgap if we observe $A_t$. A shallow valley of $A_t$ near $\omega_r$ in the far field also attributes to the self-adaptive bandgap. In the passbands of the NAM beam with $k_n$=1e9, the wave propagation approximates to the linear case.

### 3.3 Influences of the nonlinear strength on the wave propagation

The nonlinear strength influences the properties of NAM greatly [24], as the amplitude $h_1$ or the nonlinear stiffness coefficient $k_n$ determines the bifurcations. As verified [23], laws arising from changing $h_1$ and $k_n$ are same.

There is an interesting peak for THG on branch-3. As shown in Figure 12, $k_n$ influences the peak's frequency. At a given frequency, $h_3$ also features three branches under the varying $k_n$, and a saddle-node bifurcation $k_{nJ}$ connects branch-2 and branch-3. Different frequencies' branch-1&2 are approximate. Because $f_F$=75 Hz is near the cutoff frequency of the linear LR bandgap, the peak will not appear at 75 Hz no matter how strong the nonlinearity is. Slightly increasing $f_F$ to 76 Hz, a distinct peak germinates at the weak nonlinearity $k_n$=1.1e8. If $f_F$ >76 Hz, the peak always appears on branch-3 and a higher frequency requires a larger $k_n$ to generate that peak. Moreover, if $k_n$<5e7, this peak disappears, as predicted by the WN solution. Damping in the oscillators has a major influence on the amplitude of the peak but not on other parts of the branch; unexpectedly, the peak with a lower frequency obtains larger reductions, so that the peak at 76 Hz under $\zeta$=0.02 even disappears.

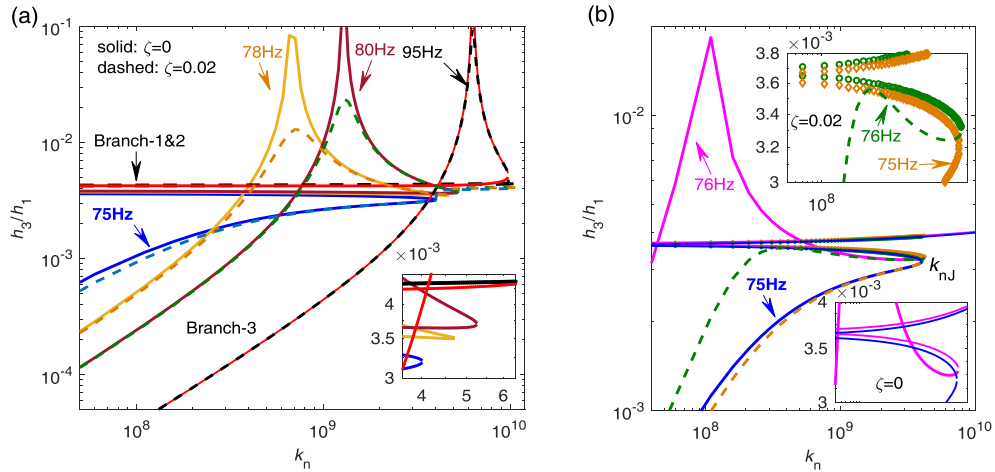

**Figure 12.** Influences of $k_n$ on $h_3/h_1$ at typical frequencies solved with UC model. Solid curves delineate undamped results. Dashed curves correspond to $\zeta$=0.02. (a) Four frequencies $f_F$=75, 78, 80 and 95Hz; (b) detailed plots of 75 and 76 Hz.

Apparently, the stronger the nonlinearity is, the higher bifurcation frequency $\omega_J$ will be for the nonlinear fundamental resonance and amplitude-frequency curves of THG, as shown in Figure 13. As predicted by the analyses above, a higher $\omega_J$ leads to a broader band for highly efficient THG inside the bandgap. Numerical results illustrated in Figure 13 demonstrate this prediction. For the case N2: $k_n$=1e10 N/m$^3$, $\omega_J>\omega_{cL}$, therefore the whole LR bandgap is full of third harmonics. The jumping frequency presented in numerical results fits UC solutions though there are errors. Moreover, $k_n$ influences little on branch-1 for $\omega_F>\omega_{cL}$. For $\omega_F<\omega_J$, a weaker nonlinearity results in the smaller $h_3$ on branch-1, including the peak at $\omega_r/3$; meanwhile, the curve near $\omega_r$ becomes steeper. However, the peak like the WN solution actually never appears, though the nonlinearity is very weak.

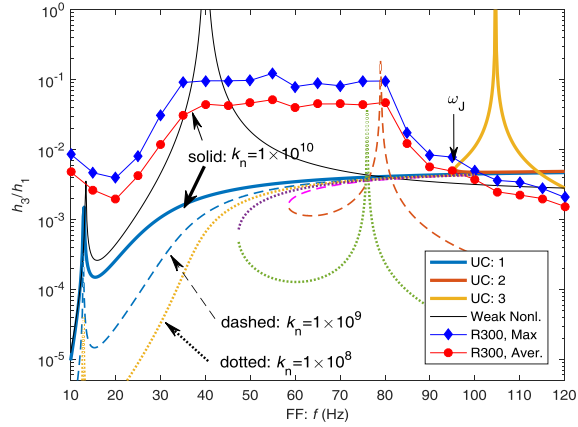

**Figure 13.** Third-harmonic amplitude $h_3$ solved with analytical and numerical approaches mainly for the strong nonlinearity $k_n$=1e10 N/m$^3$. The dashed (dotted) curves are the three branches of $k_n$=1e9 ($k_n$=1e8) solved with UC method shown.

Figure 14 shows the distributions of generalized amplitudes $A_t$, $A_1$ and $A_3$ for $k_n$=1e10 N/m$^3$. This strongly nonlinear case features highly efficient THG in 35-80 Hz. The distance- amplitude-dependent bandgap appears again. Because $\omega_J>\omega_{cL}$, this NAM beam does not have a near-field ($n<20$) bandgap for the fundamental wave in 40-73 Hz. A longer propagation distance is required to open that bandgap than the case $k_n$=1e9.

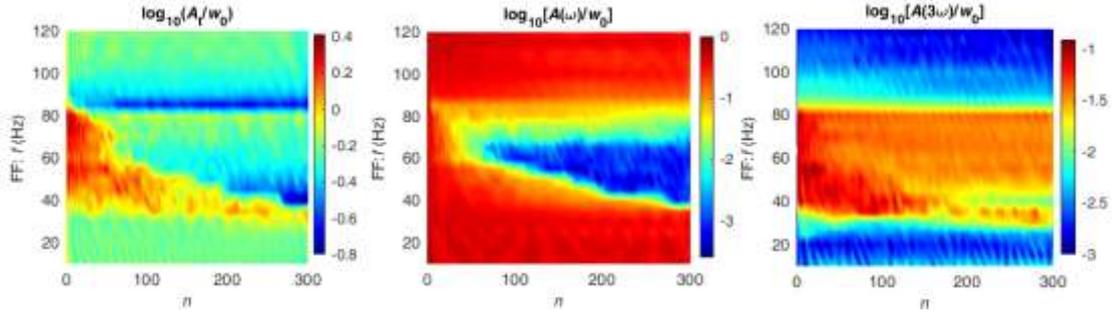

**Figure 14.** Distributions of generalized amplitudes for $k_n$=1e10.

However, the case N2 obtains obvious reductions of the fundamental waves in 75-100 Hz (the linear passband), as detailed in Figure 15. Im($k_1$) shown in Figure 16a explains its mechanism. As clarified above, the interval ($\omega_J$, $\omega_{cN}$) dominated by branch-3 is the NLR bandgap of fundamental waves, in which Im($k_1$)>0. In theory, there is still a narrow NLR bandgap in the interval 95.6< $f_F$ <102.4 Hz in the case N2 permitting the attenuation of fundamental waves in 75-100 Hz.

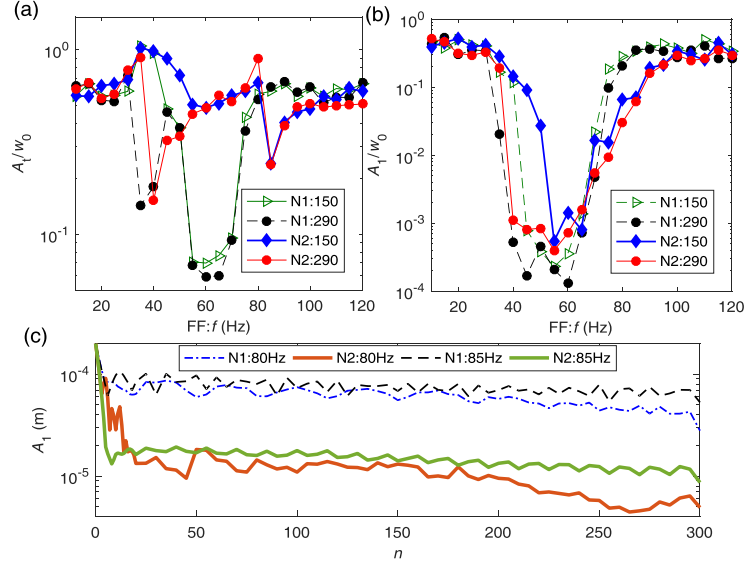

**Figure 15.** Detailed plots of wave propagations in two cases. N1: $k_n$=1e9; N2: $k_n$=1e10. (a) $A_t$ and (b) $A_1$ at the 150$^{th}$ and 290$^{th}$ cells. (c) $A_1$ varying with the propagation distance $n$ at $f_F$=80 and 85 Hz.

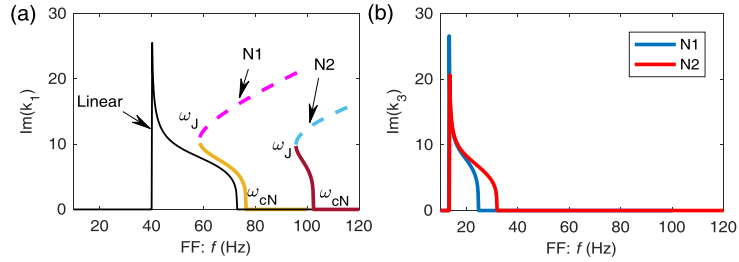

**Figure 16.** Imaginary parts of wave numbers $k_1$ and $k_3$ in two cases. N1: $k_n$=1e9; N2: $k_n$=1e10.

As shown in Figure 14, another phenomenon observed is the reduction of $A_3$ with the distance $n$ near 30~40 Hz that never happens in the case N1 (see Figure 10). This attributes to the cutoff frequency for Im($k_3$)>0 is shifted from 24.8 Hz to 32 Hz, so that the attenuation band for the third harmonic with mode $k_3$ is expanded in that frequency range.

## 4. DISCUSSIONS AND CONCLUSIONS

This work reports the wave propagation in the infinite nonlinear acoustic metamaterial (NAM) beam (consisting of periodic Duffing oscillators) by considering the third harmonic generation (THG). Characteristics and bifurcation mechanisms of the nonlinear resonance in a cell, the effective density, wave numbers, NLR bandgap, propagations and couplings of the fundamental and the third harmonics, including interrelationships between them, are elucidated.

(1) Firstly, analytical mathematical frameworks are proposed to describe the fundamental wave propagation, THG and interactions in the equivalent homogenous medium. It shows that the wave number of the fundamental wave derived by $ρ_{eff}$ by specifying the displacement (but not force) is equal to the UC solution. Finite element models demonstrate that both UC and FC solutions describe correctly the amplitude-frequency relationship of the third harmonic, including the accurate saddle-node bifurcation frequency $ω_J$.

(2) Secondly, the distribution and bifurcations of the third harmonic is clarified. In the whole

frequency range, the THG efficiency inside the linear LR bandgap is highest and the amplitude of the third harmonic $h_3$ jumps to branch-3 at $\omega_J$. Besides the peaks near the nature frequency $\omega_r$ and $\omega_r/3$, $h_3$ features an interesting peak on branch-3. This peak depends on the nonlinear strength and damping. Although the dominant wave mode of the third harmonic changes with the frequency, its amplitude almost keeps steady with the increasing propagation distance.

(3) Moreover, we find that different indexes have an identical $\omega_J$, which benefits the understanding and prediction of NAM. $\omega_J$ initiates the near-field NLR bandgap of fundamental waves, whose width is narrower for a stronger nonlinearity attributed to the slowly shifted cutoff frequency, as demonstrated by the dispersion curves of NAMs in Ref. [25]. Furthermore, it is found that the start frequency of the NLR gap decreases as the increasing of the propagation distance due to the attenuated amplitude. Therefore, the far-field NLR bandgap approaches to the original linear LR gap. Such a distance-amplitude-dependent property leads to the self-adaptive width of NLR bandgap.

(4) In addition, when $\omega_J<\omega_{cL}$, wave propagations in the passbands of linear and nonlinear AMs are similar. This law is different from the chaotic band of finite NAMs, in which the broadband resonances of the finite structure are reduced greatly [25]. While if $\omega_J>\omega_{cL}$ due to the extremely strong nonlinearity, the whole linear LR gap is full of third harmonics; the narrow near-field NLR gap still exists, which gives rise to broader reduction of fundamental waves in the far field.

In short, our work proposes effective approaches to study NAMs and unveils essential properties of the wave propagation in infinite NAMs. The clarified phenomena enable future studies and constructions of NAMs with novel properties.

## ACKNOWLEDGEMENTS

This research was funded by the National Natural Science Foundation of China (Projects Number 51405502, number 51275519).